\documentstyle[prl,multicol,epsf,epsfig,aps]{revtex}
\newcommand{\BEQ}{\begin{equation}}
\newcommand{\EEQ}{\end{equation}}
\newcommand{\BEA}{\begin{eqnarray}}
\newcommand{\EEA}{\end{eqnarray}}

\renewcommand{\d}{{\rm d}}

\renewcommand{\v}{v}

\newcommand{\half}{\frac{1}{2}}
 
\newcommand{\CH}{{\cal H}}

\def\dbarrm {{\mathchar'26\mkern-11mu{\rm d}}}                       %
\begin{document} 
\draft
\title
{A mathematical theorem as basis for the second law:
Thomson's formulation applied to equilibrium}
\date{\today}
\author{A.E. Allahverdyan$^{1)}$ and Th.M. Nieuwenhuizen$^{2)}$}
\address{
$^{1)}$Yerevan Physics Institute,
Alikhanian Brothers St. 2, Yerevan 375036, Armenia. \\
$^{2)}$ Institute for Theoretical Physics,
University of Amsterdam\\
Valckenierstraat 65, 1018 XE Amsterdam, The Netherlands. \\ 
}
\maketitle

\begin{abstract}
There are several formulations of the second law, and they may, 
in principle, have different domains of validity.
Here a simple mathematical theorem is proven which serves as the
 most general basis for the second law, namely 
the Thomson formulation (`cyclic changes cost energy'),
applied to equilibrium.
This formulation of the second law
is a property akin to particle conservation (normalization
of the wavefunction). It has been stricktly proven for a canonical ensemble,
and made plausible for a micro-canonical ensemble.

As the derivation does not assume time-inversion-invariance, 
it is applicable to situations where persistent current occur.
This clear-cut derivation allows to revive the 
``no perpetuum mobile in equilibrium'' formulation 
of the second law and to criticize some
assumptions which are widespread in literature. 

The result puts recent results devoted to foundations and 
limitations of the second law in proper perspective, and 
structurizes this relatively new field of research.
\end{abstract}

\section{Introduction}
The second law is undoubtedly one the most known statements of
statistical thermodynamics \cite{landau}. Its most known formulation is
`the entropy of a closed system cannot decrease'.
Despite of its important role  in the modern science | or may be
even {\it due} to this role | its typical formulations are frequently
folklore-minded and not very explicit. After all, what is precisely 
meant by entropy? Moreover, the law is rarely
formulated rigorously \cite{landau}. This has led to a pertinent
opinion that the second 
law is an empiric relation which is supported by observations, and
at least not inconsistent with the formalism of quantum physics. 
This situation is especially unfortunate, since the
absence of explicit formulations of the second law makes it
difficult to study its generalizations or to limit
its domain of applicability in extreme (quantum) conditions
\cite{cap,nik,AN,ANLandauer,ANperpmob}.
This became additionally complicated by the fact that the most typical
formulations of the second law use the concept of entropy,
which is a context-dependent quantity and which is 
frequently not  observed directly. Indeed, the standard definition
$\d S=\dbarrm Q/T$ is only an {\it identification} of the measured heat 
with a change in the {\it thermodynamic} entropy; 
`measuring' entropy can be done in numerics
if one determines the fraction of time that states are visited
~\cite{SKMa}, but other definitions of entropy 
occur as well~\cite{MaesLebowitz}. 
All by all this led to a disappointing
situation, where far less known and less important subjects of
statistical physics received much attention, while 
the second law itself
still keeps its not very explicit and vague look. 
The situation became acute, when we discovered that several 
formulations of the second law (Clausius inequality,
positivity of energy dispersion and entropy production)
are violated in the standard model for quantum brownian motion, which
is a harmonic quantum particle coupled to a bath of
harmonic oscillators~\cite{AN,ANperpmob}.

In the present paper we try to bridge this gap, and restate that the
second law of thermodynamics | as formulated by Thomson | is just a
rigorous theorem of quantum mechanics, comparable to 
particle conservation (normalization
of the wavefunction). Standard quantum mechanics
completely suffices for derivation of the theorem and its adequate
interpretation. The Thomson formulation | and this is its main
advantage over all other formulations | uses the unambiguous and
well-defined concept of {\it work}.
In contrast to entropy, work is a relatively straightforward 
quantity, and its use does not assume any particular caution.
The proposed clear-cut Thomson formulation will allow us to establish 
a connection between the third and second laws, to analyze certain
opinions expressed in literature about the second law, 
as well as to put into the
proper perspective the recent attempts towards identification of 
limits of validity for the second law \cite{AN}. 
From the mathematical viewpoint
the presented results are not completely new, since the main theorem
appeared with a different, more complicated proof in works  
of Pusz and Woronowicz \cite{W} and Lenard \cite{Lenard}. 
The purposes of these authors were quite different from ours, since
they used the theorem as an argument towards describing
the quantum equilibrium state through the Gibbs distribution.  

For a general, pedagogic text on the history and
today's status of thermodynamics and the second law, 
we refer to the recent work by Uffink~\cite{Uffink}.
For a collection and discussion of the original papers, see
the book by Kestin~\cite{Kestin}. A very recent discussion of the
second law within the axiomatic thermodynamics
was presented by Lieb and Yngvason~\cite{LiebYngvason}.
A dialogue on the some of  the definitions of entropy 
was reproduced by Maes and Lebowitz~\cite{MaesLebowitz}.

The setup of this paper is as follows.
In section II we will derive the theorem for the quantum 
mechanical situation and in section III we consider the derivation
for the classical case. In section IV we close with a discussion.

\section{Quantum mechanical proof of Thomson's formulation in 
equilibrium}

Here we shall present a general proof of Thomson's formulation 
of the second law as applied to equilibrium: 
{\it No work can be extracted from a closed equilibrium system during 
a cyclic variation of a parameter by an external source}.

The idea of the following derivation was given by Lenard \cite{Lenard}. 
He was adopting to the physical language a more general proof 
given in \cite{W}. This last proof is fairly difficult for the average
physically-minded reader, since it uses the techniques 
of $C^{*}$-algebras. As a by-product of our present consideration we will 
significantly simplify the original derivation of Lenard.

A closed quantum statistical system is considered. The dynamics is described
by the Hamiltonian $\CH_0$. At the moment $t=0$ an external 
time-dependent field is switched on, and the Hamiltonian becomes
${\cal H}(t)$. This field represents the influence of an external,
deterministic source.
The field is 
switched off at the moment $t$, and the 
Hamiltonian will be again $\CH_0$. Thus we have a cyclic variation of a
parameter with at least one period. 
Neither the explicit character of this parameter, nor
the Hamiltonians $\CH_0$ and ${\cal H}(t)$ have to be specified.
It is only assumed that initially, before the variation has started,
the system was in the equilibrium state described by the Gibbs
distribution: 
\begin{equation}
\label{isfahan}
\rho (0)=\frac{e^{-\beta \CH_0}}{Z}, \qquad 
Z={\rm tr}~e^{-\beta \CH_0},
\end{equation}
where $\beta=1/T$ is the positive inverse temperature.
In the time-interval $t$ the source of the external field does
work on the system. Since the system is closed before and after the
variation, the work is equal to the difference between the final and
initial energies:
\BEA
W={\rm tr}\,\{\,\CH_0\,[\,\rho (t)-\rho(0)\,]\,\}.
\EEA
It can also be written alternatively as
\begin{equation}
\label{work}
W=\int^{t}_{0}\d s~{\rm tr}[\rho (s)\frac{\d {\cal H}(s)}{\d s}],
\end{equation}
where one uses integration by parts, and the equation of motion:
\begin{equation}
\label{motion}
i\hbar\frac{\d\rho (t)}{\d t}=[{\cal H}(t),\rho (t)].
\end{equation}
Let us now go to the interaction representation and introduce 
a unitary operator $V$ as
\begin{equation}
\label{7}
\rho (t)=e^{it\CH_0/\hbar}\,V\,\rho (0)\,V^{\dagger}\,e^{-it\CH_0/\hbar}.
\end{equation}
Eq.~(\ref{work}) now reads
\begin{equation}
\label{8}
W={\rm tr}[\,\CH_0\,V\,\rho (0)\,V^{\dagger}\,]-{\rm tr}[\CH_0\,\rho (0)].
\end{equation}
It is seen that as far as the work is concerned, any cyclic 
variation enters only through its corresponding unitary operator $V$. Our aim
now is to show that $W$ defined by (\ref{8}) is nonnegative, or in
other words, the final average energy is not smaller than the initial
one. Notice especially that we compare only the average energies.

Due to Eq.~(\ref{isfahan}) $\rho (0)$ and $\CH_0$ commute, and 
thus have a common eigenbasis $|k\rangle$. 
Let us denote eigenvalues of $\rho (0)$ as $\{ r_k\}$,
and those of $\CH_0$ as $\{ h_k\}$. It holds that $r_k=\exp(-\beta h_k)/Z$. 
For simplicity we will consider a finite dimensional Hilbert space.
One has
\begin{equation}
\label{37}
{\rm tr}[\CH_0\,V\rho (0)\,V^{\dagger}]=\sum_{m,k=1}^n h_m\v_{mk}r_k,
\end{equation}
where $n$ is the dimension of the corresponding Hilbert space, and 
$\v_{mk}=\langle m|V|k\rangle\langle k|V^{\dagger}|m\rangle$.
Since $V$ is unitary, $VV^{\dagger}=V^{\dagger}V=1$, it follows that 
$\v_{mk}$ is double-stochastic:
\begin{equation}
\label{39}
\v_{mk}\ge 0,\qquad 
\sum_{m=1}^n \v_{mk}=\sum_{k=1}^n \v_{mk}=1.
\end{equation}
One arranges the $h_m$ in a non-decreasing order 
\begin{equation}
\label{43}
h_1\le h_2\le .....\le h_n,
\end{equation}
which implies (due to the fact that the exponential function is monotonic) 
that  the $r_m$ are arranged as
\begin{equation}
\label{41}
r_1\ge r_2\ge .....\ge r_n\ge 0.
\end{equation}
The work (\ref{8}) reads in these variables
\begin{equation}
\label{77}
W=\sum_{m=1}^n h_ms_m-\sum_{m=1}^n h_mr_m
\end{equation}
where $s_m$ is defined as
\begin{equation}\label{12}
s_m=\sum_{k=1}^n \v_{mk}r_k.
\end{equation}
Now we employ a summation by parts (the discrete analog
of integration by parts)
\BEQ
\sum_{m=1}^n h_ms_m  =
-\sum_{m=1}^{n-1}(h_{m+1}-h_m)\sum_{i=1}^m s_i
+h_n\sum_{k=1}^ns_k,
\EEQ
and the same with $r_m$ replacing $s_m$, to obtain 
\BEQ \label{772}
W=\sum_{m=1}^{n-1}(h_{m+1}-h_m)\sum_{i=1}^m (r_i-s_i)
+h_n\sum_{k=1}^n(s_k-r_k)=\sum_{m=1}^{n-1}(h_{m+1}-h_m)\sum_{i=1}^m (r_i-s_i)
\end{equation}
In the second step Eq. (\ref{39}) was used.
To prove that $W\ge 0$, notice that $h_{m+1}-h_m\ge 0$. Therefore it suffices to 
show that
\begin{equation}
\label{45}
\sum_{i=1}^{m}r_i\ge \sum_{i=1}^{m}s_i,
\end{equation}
Hereto one denotes
$\phi_{k}^{(m)}=\sum_{j=1}^m \v_{jk}$, which has the properties
\BEA
0\le \phi_{k}^{(m)}\le 1,\qquad \sum_{k=1}^n\phi_k^{(m)}=m,
\label{141}
\EEA
as follows from (\ref{39}). One then gets
\BEA
\label{rashid12}
\sum_{i=1}^m(r_i-s_i)&&=\sum_{i=1}^mr_i-\sum_{k=1}^n\phi_k^{(m)}r_k
=\sum_{k=1}^m(1-\phi_k^{(m)})r_k-\sum_{k=m+1}^n\phi_k^{(m)}r_k \EEA
Now using the ordering (\ref{41}) of the $r_k$, one gets the lower bound
\BEA
\label{rashid1}
\sum_{i=1}^m(r_i-s_i)&&\ge
\sum_{k=1}^m(1-\phi_k^{(m)})r_m-\sum_{k=m+1}^n\phi_k^{(m)}r_m 
=\left(m-\sum_{k=1}^n\phi_k^{(m)}\right)r_{m}=0,
\label{rashid2}
\EEA
where the last step follows because of Eq. (\ref{141}). 
Therefore Eq. (\ref{45}) has now been proven. Inserting this in Eq. (\ref{772})
one finally has
\begin{equation}
\label{79}
W\ge 0
\end{equation}

This derivation concludes the proof of Thomson's formulation of the second law 
for this case: from a system in the equilibrium state work cannot be extracted in
a cyclic process. 
The inequality sign  says that work can nevertheless be done 
on the system, as is physically obvious.

At zero temperature the equilibrium state is the ground state.
The inequality $W\ge 0$ is then obvious without any derivation, and
confirms that no work can be extracted from the 
ground state.

\section{Microcanonical initial distribution}

The above analysis assumes that the initial state of the system 
is the Gibbs distribution. 
In the present section we shortly consider some other distributions
of statistical physics. First let us notice that we only used 
two properties of the Gibbs distribution: commutation with the initial
Hamiltonian and the opposite ordering of the corresponding
eigenvalues, as given by Eqs.~(\ref{43}, \ref{41}). 
Thus, the no work-extraction principle: $W\ge 0$, is valid for all
initial distributions which satisfy these properties. As a
particular case, we mention the generalized microcanonical distribution
or $\theta$-distribution \cite{st} 
\BEA
r_k=\frac{1}{m}\,\theta(m-k), \qquad 1\le k\le n, \qquad 1<m<n,
\EEA 
where $\theta(k)$ is step function: $\theta(x\ge 0)=1$,
$\theta(x<0)=0$. Thus all energy levels below a fixed level $h_m$ are
equally populated, while the energy levels larger than $h_m$ are not
populated. The monotonicity properties (\ref{43}, \ref{41}) are
obviously satisfied, so that $W\ge 0$ is also valid for the present case.

For the strictly micro-canonical ensemble one considers an energy
shell $({\cal E-\d{\cal E}}, {\cal E})$,
which is a group of energy levels such that the 
difference ${\d\cal E}$ between the maximal and the minimal energy level of the
shell is smaller than a characteristic uncertainty of energy 
\cite{landau}. Let the total number of levels within the shell be
$\Omega$. The states within the shell ${\d\cal E}$ are equally probable,
\BEA r_k=\frac{1}{\Omega}
\EEA
for any level $h_k$ belonging to the shell, while for other levels one has $r_k=0$. 
It is seen that the arrangement of  $r_k$'s is non-monotonous as soon as 
the shell is located above the vacuum, i.e. if the minimal energy of the shell
is higher than the vacuum energy. For this case a straightforward application 
of the above theorem is impossible. Let us see where precisely our proof fails.
The simplest situation of this kind is a shell which consists of one single 
non-vacuum energy level.  For simplicity suppose that we are trying to check 
Eq.~(\ref{45}) with a distribution $r_1=0$, $r_2=r_3....= r_n>0$ 
(this is a shell with $n-1$ levels, which just starts one level
above the vacuum). As expected,  Eq.~(\ref{45}) can be violated for $m=1$ 
\BEA
\label{cc}
r_1-s_1=-\sum_{k=2}^n\phi_k^{(1)}r_k\le 0.
\EEA
More generally, a negative contribution to the work arises from systems
that, due to the cyclic process, end up in energy levels below the shell. 
As a result, the theorem cannot hold in full generality, and may not apply, 
e.g., to small microcanonical systems.

The above arguments are simple enough to convince us that
a proper formulation of the second law for the microcanonical
ensemble should be connected with certain limitations. 
As already discussed, one way is to  require that the shell is so 
wide that it includes the vacuum state, and then one has the 
$\theta$-distribution or generalized microcanonical distribution; 
the validity of $W\ge 0$  was shown above.
Another way is to consider only those unitary operations which do not bring
system to energies less than the lower shell-limit ${\cal E}-\d{\cal E}$; 
under this condition the theorem again applies, since the dangerous terms 
(those with energies below the shell) have vanishing matrix element $\v_{mk}$. 
For large systems it is well known that almost all states are very near the
maximal limit ${\cal E}$ of the shell. Let us suppose that for a macroscopic
system with $N$ degrees of freedom, ${\cal E}$ is proportional to $N$. 
The shell thickness $\d{\cal E}$ has to be much smaller than the typical uncertainty 
$\sqrt{N}$ of the energy. Let us choose $\d{\cal E}\sim N^\alpha$ with $\alpha<\half$.
Now given the fact that almost all systems of the ensemble have energy very close
to the upper bound ${\cal E}$, extraction of energy less than $N^\alpha$ 
is ruled out for almost all members of the ensemble. 
On physical grounds one then expects that extraction of more energy is also very 
unlikely. This means that the central inequality $W\ge 0$ holds
for all practical purposes in the micro-canonical ensemble.

\section{Classical proof of Thomson's formulation.}

The above quantum result remains valid if the spectrum becomes dense, so the
classical case is included into the consideration. Nevertheless, for
the interested reader we will briefly outline the proof of the second
law, when starting immediately from the classical formalism. Here the
state of the system is described through the probability density
${\cal P}(p,q,t)$ as a function of time $t$, the canonical momentum
$p$, and the canonical coordinate $q$ (in fact $p$ and $q$ can be
arbitrary dimensional vectors; since the generalization to this case
is obvious, it will not be discussed by us separately). The initial
Hamiltonian and the time-dependent Hamiltonian are still denoted by
$\CH_0$ and ${\cal H}(t)$, respectively. The evolution of ${\cal P}$ 
is described by the Liouville equation: 
\BEA
\label{rusha}
\partial_t{\cal P}(t)={\cal L}(t)\,{\cal P}(t),\qquad {\cal P}(t)
={\cal T}e^{\int_0^t\d s\,{\cal L}(s)}{\cal P}(0),
\EEA
\BEA
{\cal L}(t)
\equiv \frac{\partial{\cal H}(p,q;t)}{\partial q}\,
\frac{\partial }{\partial
p}-\frac{\partial{\cal H}(p,q;t)}{\partial p}\,
\frac{\partial }{\partial q},\qquad
{\cal T}e^{\int_0^t\d s\,{\cal L}(s)}\equiv
1+\sum_{k=1}^{\infty}\int_0^t\d s_1\int_{s_1}^t\d s_2...
\int_{s_{k-1}}^t\d s_k {\cal L}(s_k)...{\cal L}(s_1), 
\label{eraz}
\EEA
where ${\cal L}$ is the Liouville operator and ${\cal T}$ is the
chronological symbol as defined above. For a cyclic variation
the work in the classical situation reads:
\BEA
W=\int\d p\,\d q\,\CH_0(p,q)\,[\,{\cal P}(p,q,t)-{\cal P}(p,q,0)\,]. 
\EEA
Eq.~(\ref{rusha}) can also be
written in the integral form:
\BEA
{\cal P}(p,q,t)=\int\d p'\,\d q'\,{\cal P}(p',q',0)\,{\cal
K}(p,q;t|p',q';0),\qquad {\cal K}(p,q;t|p',q';0)=
{\cal T}e^{\int_0^t\d s\,{\cal L}(s)}\,
\delta(p-p')\,\delta(q-q').
\label{ggg}
\EEA
Now the work done by the external source reads:
\BEA
\label{korund}
W=\int\d p\,\d q\,\d p'\,\d q'\,\CH_0(p,q)\,{\cal P}(p',q',0)\,
{\cal K}(p,q;t|p',q';0)-\int\d p\,\d q\,\CH_0(p,q)\,{\cal P}(p,q,0).
\EEA
The analogy with Eqs.~(\ref{8}, \ref{37}) is by now fully obvious. In
particular, the role of the discrete indexes $i$ and $k$ in those
equations is now played by the continuous double-indices $(p,q)$ and
$(p',q')$; the role of $v_{ik}$ is played by ${\cal K}(p,q;t|p',q';0)$. 
Due to its definition (\ref{ggg}), ${\cal K}(p,q;t|p',q';0)$ 
does have the standard properties of the conditional probability 
distribution, and one additional property which makes it a
double-stochastic (continuous) matrix: 
\BEA
{\cal K}(p,q;t|p',q';0)\ge 0,\qquad \int\d p\,\d q\,{\cal
K}(p,q;t|p',q';0)=\int\d p'\,\d q'\,{\cal
K}(p,q;t|p',q';0)=1.
\EEA
The only non-trivial property is the last one, but it quite clearly
follows from (\ref{eraz}, \ref{ggg}) upon noting that ${\cal L}$ is a
differential operator and integrals similar to $\int \d p'\,\d q'\,
{\cal L}(s_1)...{\cal L}(s_k)\delta(p-p')\,\delta(q-q')$ 
are equal to zero.

Once the property of the double-stochasticity and the essential
similarity between (\ref{8}, \ref{37}) and (\ref{korund}) is
established, it is a matter of repetition to derive the proof of 
\BEA
W\ge 0
\EEA 
in the classical case. The reader should notice that by saying this
we ignore all convergence problems which can arise due to the
continuous character of the considered classical situation. 
The most reasonable
way to overcome such problems is to introduce an additional
regularization. However, we will leave the situation
as it is, since the readers who are sensitive to this kind of problems
are just invited to get the classical situation as the limiting case
of the above quantum proof (after all, the quantum formulation
 is the physical way to regularize the classical problem).

\section{Discussion}

In classical physics there are many equivalent formulations 
of the second law. Examples are: non-decrease of entropy of a closed system,
heat goes from high temperature to low temperature, 
the Clausius inequality $\dbarrm Q\le T\d S$ , non-negativity of the rate of
entropy production and  non-negativity of the rate of energy dispersion.
A more folkore minded example is the absence of steady currents.
In recent studies of quantum systems, several of these formulations 
have been questioned~\cite{cap,nik,AN,ANperpmob,ANLandauer}.
The fundamental question is then whether there is still a unique
formulation that is satisfied in all cases. 
The aim of this paper is to demonstrate that there
indeed exists such a formulation, and it is related with work,
which, fortunately, is more accessible than heat and certainly more accessible
than entropy, for which there are many definitions~\cite{MaesLebowitz}.

In the present section we will discuss the above theorem and its
relations with the standard understanding of the second law, as well
as we outline some relatively straightforward applications of the
above theorem.

It is clear that the theorem forbids even one single work extraction
cycle. This is to be put in contrast with the following known 
version of the second law: No {\it perpetuum mobile of the second kind}
(i.e., a device which makes as many work-extracting cycles as one
pleases) exists. It is a particular case of our theorem,
but we now point out that it is in fact much more weaker, 
as its validity depends only on the {\it existence} of a ground state. 
This ground state does not have even have to be unique, as
required for the validity of the third law \cite{landau}. Indeed,
starting from {\it any} state and making sufficiently many work-extracting
cycles with a finite extracted work per cycle, one will decrease the final average
energy of the system below its ground state energy, which is
impossible. So already a clear-cut formulation of the statement allows us
to unmask the above {\it no perpetuum mobile} statement as a
basically trivial consequence of quantum mechanics, 
rather than a deep theorem on (quantum) statistical physics.
Our new theorem (even one cycle is forbidden) heals the problem,
by forbidding `perpetuum mobile' with any finite number of 
cycles, at least as long as one starts in equilibrium.
 
When proving the above theorem we did not use any special property of
the initial Gibbs distribution, except for its commutation with the initial
Hamiltonian and the opposite ordering of the corresponding
eigenvalues (see Eqs.~(\ref{43}) and (\ref{41})). 
If the initial distribution is Gibbsian but the temperature is negative, 
this ordering property is lost, 
and the theorem does not hold. This explains the role
played by negative temperature for lasers and masers \cite{l}, where positive
work extraction is the main state of affairs.
However, for other initial distributions that do satisfy these properties, the
derivation applies as well. The most interesting case is 
the generalized micro-canonical ensemble or $\theta$-ensemble, 
where all states below a given energy are equally probable. 
In typical situations a vast majority of the states have energy very close 
to the maximal energy, implying that, at least in statics, this generalized 
ensemble is equivalent to the micro-canonical ensemble itself.  
The same property puts forward that also our theorem applies for 
all practical purposes to the micro-canonical ensemble.

Yet another line of generalization arises when one is noting that the
features of the Hamiltonian $\CH_0$ under time-inversion were
irrelevant for the proof. Thus, the studied system may well contain an
external magnetic field. In such a situation the system can contain
persistent currents in the equilibrium
state. Examples are Landau diamagnetism \cite{landau}, vortices in
conventional superconductors \cite{tin}, that may last days, boundary 
currents in the quantum Hall effect and persistent currents
in mesoscopic rings. These effects are
pretty counterintuitive from the classical thermodynamical
viewpoint, and at the first glance may even appear as a violation of 
the second law. In particular,
one of the widespread folkore-minded formulations of the 
second law refers to the impossibility
of ongoing motion in the equilibrium state, and the persistent
currents give an example of such a motion. 
Nevertheless, it does not imply any contradiction with the second law in 
the Thomson formulation, and also shows that for the present case
the time-inversion invariance does not have any direct connection with
the second law. Notice in this context that the second law in the
equilibrium Thomson formulation was proven under the condition of
time-inversion-invariance (see \cite{ru,ku} and refs. therein). 
Since the invariance property is rather strong, the authors of these
works got somewhat more detailed results than just the non-negativity of
the work $W\ge 0$. Whether these results are valid for the considered
more general case is still an open problem. 

We like to stress that our theorem also applies when the total closed system
consists of a subsystem and a heat bath, that interact with each other.
In that situation the typical case is that a work cycle is made by 
manipulating a parameter of the subsytem. This is the situation
considered in Ref. ~\cite{ANperpmob}, and it could  be checked that,
when starting from equilibrium,  the total work for making a cyclic 
change is always positive.

Let us notice that, after one cycle has been made, the system
can locally return to equilibrium. Then surplus of energy runs away 
(dissipates) in the bath.  When this process has settled, 
additional cycles cost additional work. 
Employing a standard argument, we can now show that 
non-cyclic changes, that are made in such a manner that   
afterwards one waits long enough to erase memory effects, 
also disperse energy. Indeed, by closing the cycle, there should
always be dispersion, and this is only possible
in all cases if each part disperses energy.

Finally, we would like to analyze two widespread opinions about the
second law. In their book \cite{landau} Landau and Lifshitz state
that the second law is incompatible with the microscopically
reversible quantum dynamics, and that the second law can somehow be
connected with the quantum measurement process, which in view of these
authors is an inherently irreversible process imposed on the
reversible quantum formalism. As we see above, no quantum measurement
process is directly involved into the derivation of the second law,
and the standard quantum-mechanical formalism is completely enough. 
Moreover, the dynamics of the system is unitary, i.e. it is invertible
as precisely as one wishes, so that no arrow of time is involved in the
presented derivation of the second law.

Within another school of thinking, Zurek and his
coworker \cite{z} claim that the second law does arise as a
consequence of the interaction between a quantum system and its
{\it thermal} environment (environment-induced superselection rules). 
This is again not supported by the above proof,
since it does not suppose the existence of a thermal environment,
although such a case is not excluded, provided that the
system and its environment are considered within one closed system.
Of course, this remark does not mean that the second law has nothing
to do with thermal environments in general. They are just not
necessary for the rigorous statement of the Thomson formulation
applied to the equilibrium state of a closed system. 

In conclusion, we have analyzed a mathematical theorem which serves as
a basis for the derivation of the second law in the Thomson
formulation. Once this clear-cut derivation is given, it is a matter of a
simple logic to rule out some pertinent pre-supposes on the second
law. In particular, we analyzed 
the ``no perpetuum mobile'' principle, which within the quantum theory 
was seen to be almost a trivial statement akin (and even weaker) to
the third law. 
It is hoped that the present paper will put into the proper
perspective the research devoted to the microscopical foundations 
and limitations of the second law \cite{cap,nik,AN,ANLandauer,ANperpmob}, 
since it is absolutely necessary to have a rigorous formulation of this law 
within the quantum statistical thermodynamics before
consideration of its limits and its generalizations. From our
viewpoint, neither the folklore-minded statements typically encountered in
textbooks, nor rigorous derivations within the axiomatic (formal)
thermodynamics fully meet this goal.

\section*{Acknowledgments} A.E. A. thanks Alexey Nikulov for
useful discussion, and Th.M. N. acknowledges discussion with Leendert Suttorp. 

We acknowlegde Chris Jarzynski and  Georg Reents for pointing at a 
redundant assumption  made in the early draft of this manuscript and 
Dieter Gross for drawing our attention to the microcanonical case.

\end{document}